\documentclass[a4paper,11pt]{article}

\usepackage{pos}
\usepackage{lipsum} 
\usepackage[mathlines]{lineno}
\usepackage{ragged2e}
\title{Measurement of the Three-Flavor Composition of Astrophysical Neutrinos with Contained IceCube Events}

\ShortTitle{Measurement of the Three-Flavor Composition of Astrophysical Neutrinos}

\author{The IceCube Collaboration \\{\normalsize \normalfont(a complete list of authors can be found at the end of the proceedings)}\\}

\emailAdd{abalagopalv@icecube.wisc.edu}
\emailAdd{vbasu@icecube.wisc.edu}
\emailAdd{karle@icecube.wisc.edu}

\abstract{

The IceCube Neutrino Observatory at the South Pole detects neutrinos from the entire sky, both of astrophysical and atmospheric origin, via the Cherenkov light emitted when these neutrinos interact in the ice, giving rise to rapidly moving charged particles. Neutrino events with vertices contained within the detector volume are useful for studying the neutrino flavor ratio, as they allow for a better reconstruction of the event morphology. The Medium Energy Starting Events (MESE) data sample is a selection of such events with energies of at least 1 TeV. This sample includes electron-, muon-, and tau-neutrino events, processed consistently. We use it to constrain the flavor ratio of astrophysical neutrinos at Earth, which in turn informs us of the flavor composition at the source itself. In this talk, we will present the results of this study, based on 11.4 years of IceCube data.

\vspace{4mm}

{\bfseries Corresponding authors:}
Aswathi Balagopal V.$^{1*}$, 
Vedant Basu$^{2}$, 
Albrecht Karle$^{3}$\\
{$^{1}$ \itshape University of Delaware}\\
{$^{2}$ \itshape University of Utah}\\
{$^{3}$ \itshape University of Wisconsin-Madison}\\[4mm]
$^*$ Presenter
}

\ConferenceLogo{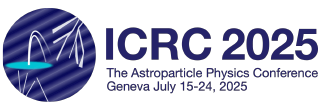}

\FullConference{39th International Cosmic Ray Conference (ICRC2025)\\
 15–24 July 2025\\
Geneva, Switzerland\\}

\begin{document}

\maketitle

\section{Introduction}\label{sec1}
 The IceCube Neutrino Observatory, located at the South Pole, detects neutrinos via the Cherenkov light deposited by their charged secondaries produced as they interact in ice~\cite{Aartsen:2016nxy}. These can be charged current (CC) or neutral current (NC) interactions of electron, muon, or tau neutrinos. 
 The Cherenkov light produced by the secondaries is detected using 5160 digital optical modules (DOMs) hosted on 86 strings inserted into the glacier. The detector measures signal from neutrinos of atmospheric and astrophysical origin with energies of hundreds of GeV and above. Muons generated from cosmic-ray air showers are also seen in the detector, and constitute the major background for astrophysical neutrino detection.

 Here we use a selection of events called 'Medium Energy Starting Events' (MESE)~\cite{MESE_ICRC}, with a total of 9888 events with interaction vertices inside the detector ("starting events") and with energies of 1 TeV and above. The selection is based on a series of veto criteria to reject muon events to obtain a neutrino-rich dataset. The MESE sample consists of electron, muon and tau neutrinos from the whole sky. Based on the distribution of light in the detector, the morphology of the observed events can be classified into cascades, tracks, and double cascades. Cascades arise from CC interactions of electron neutrinos and NC interactions of all neutrino flavors. Tracks are produced in CC interactions of muon neutrinos or tau neutrinos where the tau lepton decays into muons while other CC interactions of tau neutrinos produce double cascades. The MESE sample is also used for measuring the cosmic neutrino spectrum, as described in~\cite{MESEdiffuse:2025icrc}.
 The sample used here has a few additional events compared to that used in~\cite{MESEdiffuse:2025icrc} since an additional cut on the required number of active DOMs is implemented in~\cite{MESEdiffuse:2025icrc} and not here.
 
 Since the MESE sample has neutrinos of all flavors, it is suitable for measuring the astrophysical flavor ratio of neutrinos. Cascades and tracks within the dataset are classified as such during the selection procedure, using a deep neural network~\cite{DNN_TheoGlauch}, with $\sim 88\%$ efficiency for true cascades and $\sim 97\%$ for true tracks above 1 TeV. 
 A majority of the true double cascade events in the sample are classified as single cascades by the deep neural network, which makes it difficult to break the degeneracy between electron and tau neutrinos while measuring the flavor ratio. We therefore introduce an additional classification strategy in this analysis, designed specifically to select double cascade events. 

 \section{Method}\label{sec2}
The double-cascade selection method is likelihood-based and similar to that described in~\cite{IceCube:2020fpi}. 
Final-level events in the MESE sample, that are already classified as either cascades or tracks, are passed through a reconstruction algorithm that maximizes the likelihood under a double-cascades hypothesis, using the spatial and timing information of the deposited charge in the DOMs. 
Several variables are extracted from the event based on these reconstructions: the energy of each cascade in the double-cascade event, the sum of the energies of the two cascades, the sum of the energies deposited within 40\,m of each cascade vertex compared to the total energy, the relative energy asymmetry between the two cascades, and the decay length of the tau lepton obtained from the separation between the two cascades. 
Events that fall within a given range for these observables, predefined using simulations of neutrinos of all flavors, are retained for the double cascade classification~\cite{IceCube:2020fpi}. The final list of events accepted into the double cascade classification are events that pass these conditions, and have a reconstructed total energy greater than 30 TeV and reconstructed length greater than 10 meters to prevent misclassification from a majority of the background. These restrictions result in an expected number of 7 events, with 70\% purity. The remaining events retain their original classification as cascades or tracks.
The low count of double-cascade events is attributed to the difficulty in separating them out due to the spacing of the DOMs in the detector. A majority of tau neutrinos in the sample have energies on the TeV scale, due to the falling nature of the flux, and therefore have tau decay lengths in the scale of a few meters to a few tens of meters. This makes the separation between the two cascades difficult as the spacing between the strings holding the DOMs is $\sim$ 125 meters. 

We use observables from these classified events, both from data and simulations, to perform a forward-folded fit and measure the astrophysical flavor ratio. 
We generate 2D histograms of reconstructed energy vs. cosine of the reconstructed zenith angle for cascades and tracks. Double cascades are additionally  also binned according to the reconstructed length, to form 3D histograms. The forward-folded fit includes components that account for the conventional neutrino flux arising from pion and kaon decays in cosmic ray air showers, prompt neutrino flux from charmed decays in air showers, atmospheric muon flux, and the astrophysical neutrino flux.

We assume a broken power law (BPL) as the baseline shape of the astrophysical flux, following the best-fit measurement of the spectral shape using the MESE sample~\cite{MESEdiffuse:2025icrc}. Other IceCube measurements~\cite{CF:2023icrc} also indicate a possible break in the spectrum of cosmic neutrinos.
The physics parameters of the analysis reported here are the flavor ratios. We fit for the fraction of electron and tau neutrinos in the total astrophysical neutrino flux, and constrain the fraction of muon neutrinos via the relation $f_e + f_{\tau} + f_{\mu} = 1$. 
Systematic uncertainties in the atmospheric neutrino flux arising from uncertainties in the cosmic ray flux and the corresponding production of neutrinos are included as nuisance parameters in the fit. We also account for detector-related systematic uncertainties as nuisance parameters in the fit. These parameters allow for changes in the detector's light acceptance resulting from uncertainties in the ice model and the optical efficiency of the DOMs. 
Along with the baseline BPL fit, we also fit for a single power law (SPL) astrophysical flux assumption as a cross-check, since previous IceCube measurements were consistent with an SPL shape for the astrophysical flux. This fit is performed under the assumption that the flavor ratio remains the same across the energy scales considered here i.e. 1\,TeV to 10\,PeV. We use the software package \textsc{NNMFit} to perform the fit~\cite{NNMFit}. 

\section{Results}\label{sec2}
The number of events in the MESE sample used for the flavor measurement with 11.4 years of IceCube data is shown in Table~\ref{tab1}. This is compared to the expected number of events from the best-fit to the BPL and SPL models.

\begin{table}[h!]
\centering
\begin{tabular}{lccccccccccc}
\hline
Morphology  &	Data & BPL & SPL\\
\hline
Cascades & 4960 & $4953.6\pm 154.6$ & $4999.2\pm 160.4$ \\
Tracks & 4919 & $4876.2\pm 136.1$ & $4825.4\pm 141.7$ \\
Double Cascades & 9 & $7.0\pm 0.9$ & $9.1\pm 1.0$ \\
\hline
\end{tabular}
\caption{Number of cascades ($E>1$\,TeV), tracks ($E>1$\,TeV), and double cascades ($E>30$\,TeV) in data compared to expectation from best-fit to BPL and SPL models. Both statistical and systematic uncertainties are included in the reported errors.}\label{tab1}
\end{table}

\begin{figure}[h!]
\includegraphics[width=0.33\linewidth]{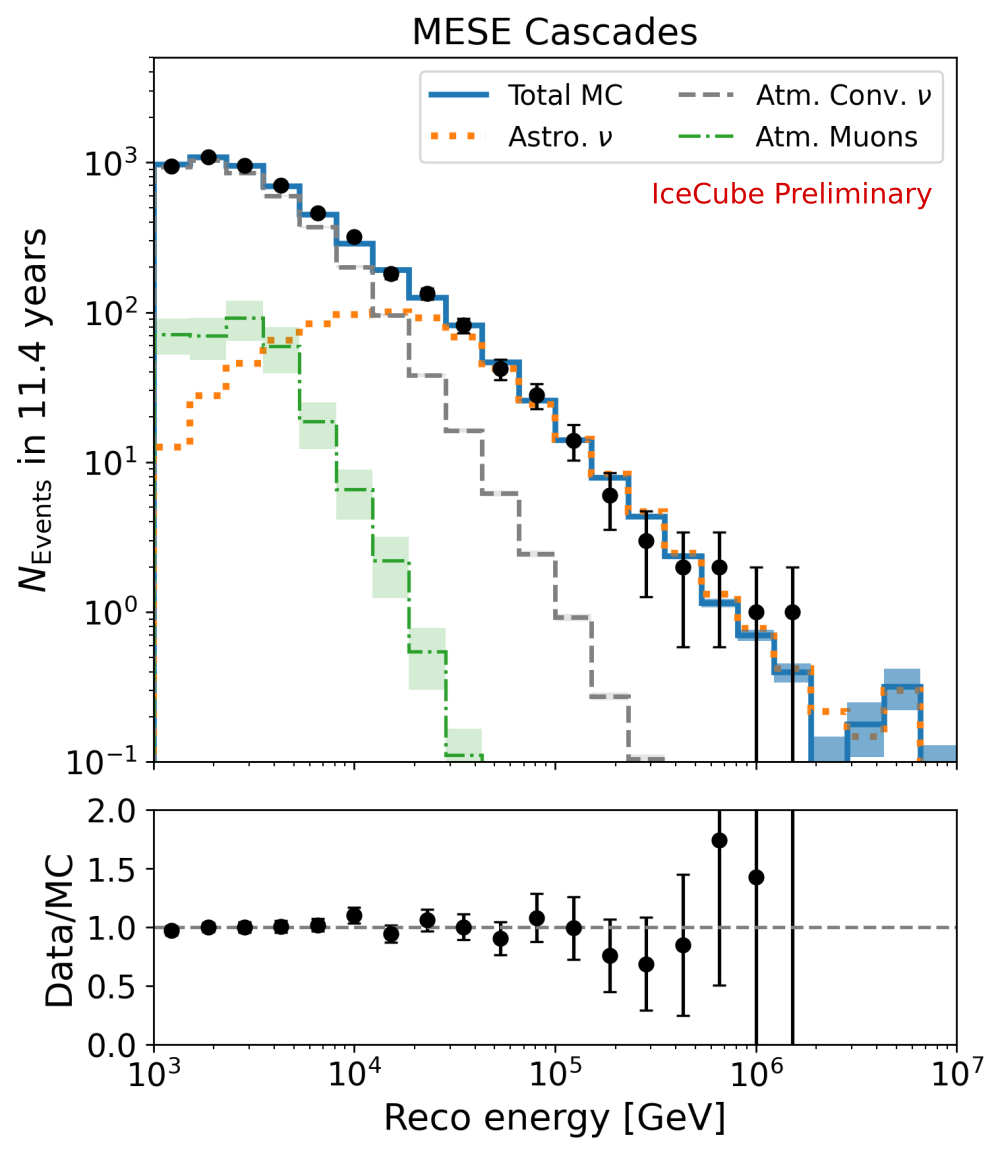}
\includegraphics[width=0.33\linewidth]{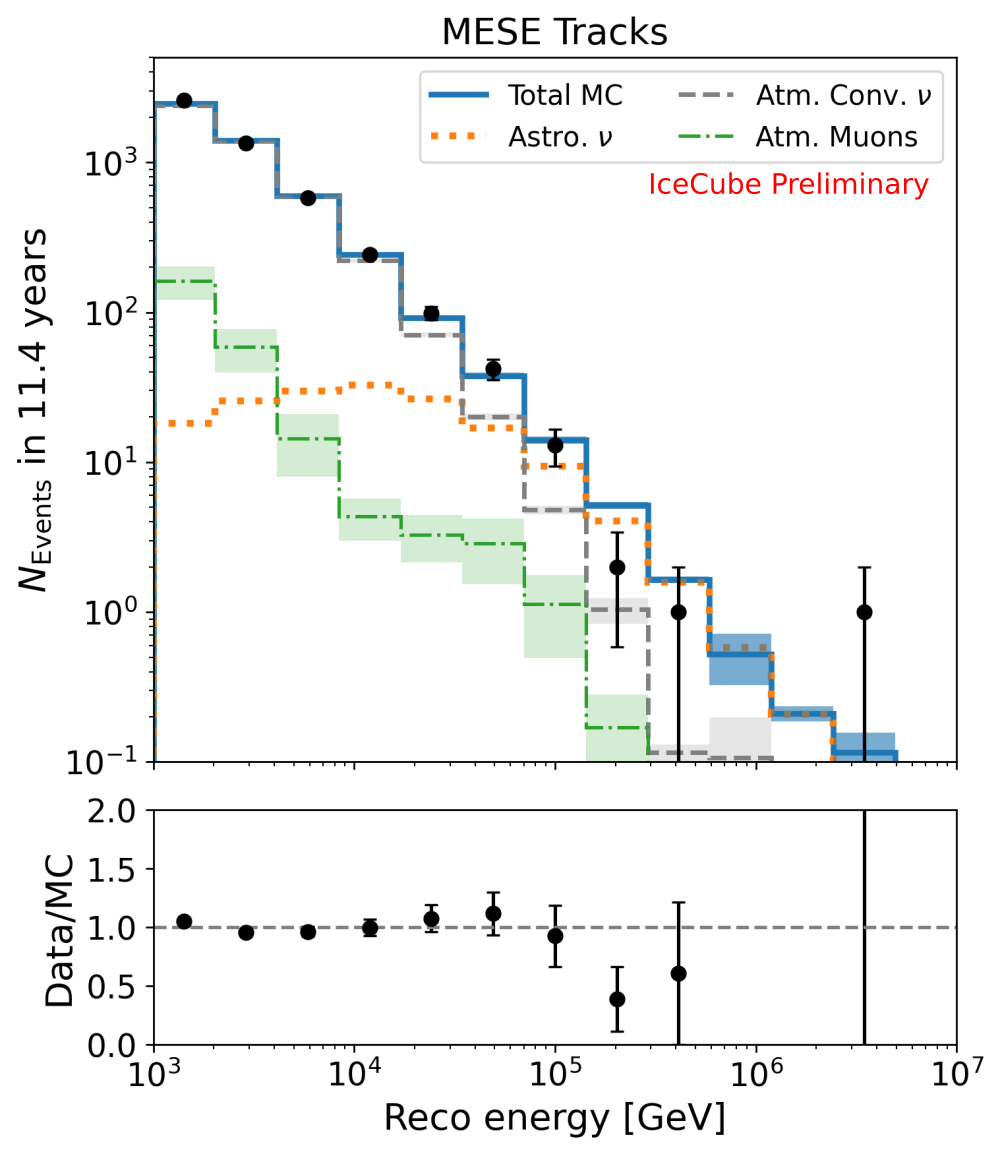}
\includegraphics[width=0.33\linewidth]{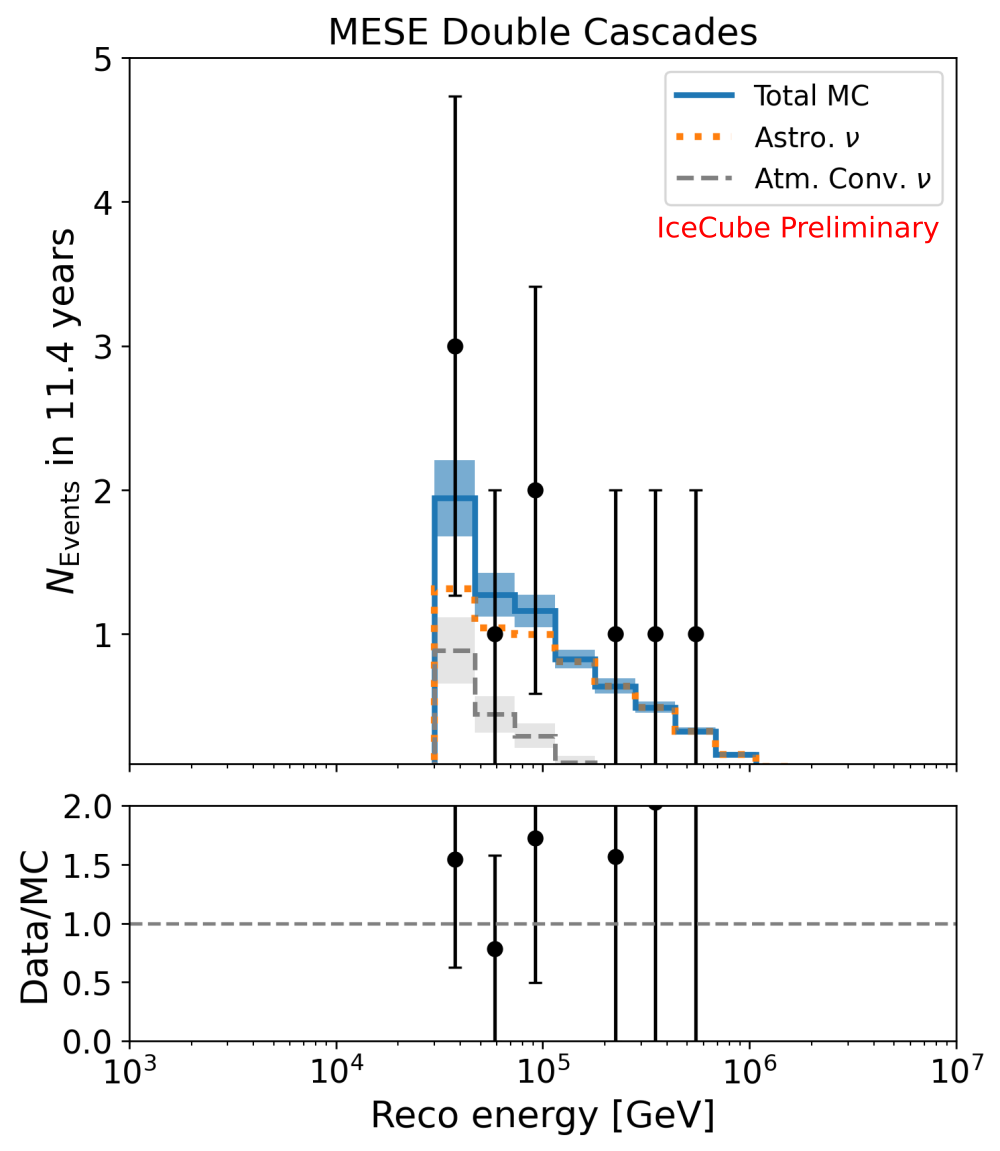}
\includegraphics[width=0.33\linewidth]{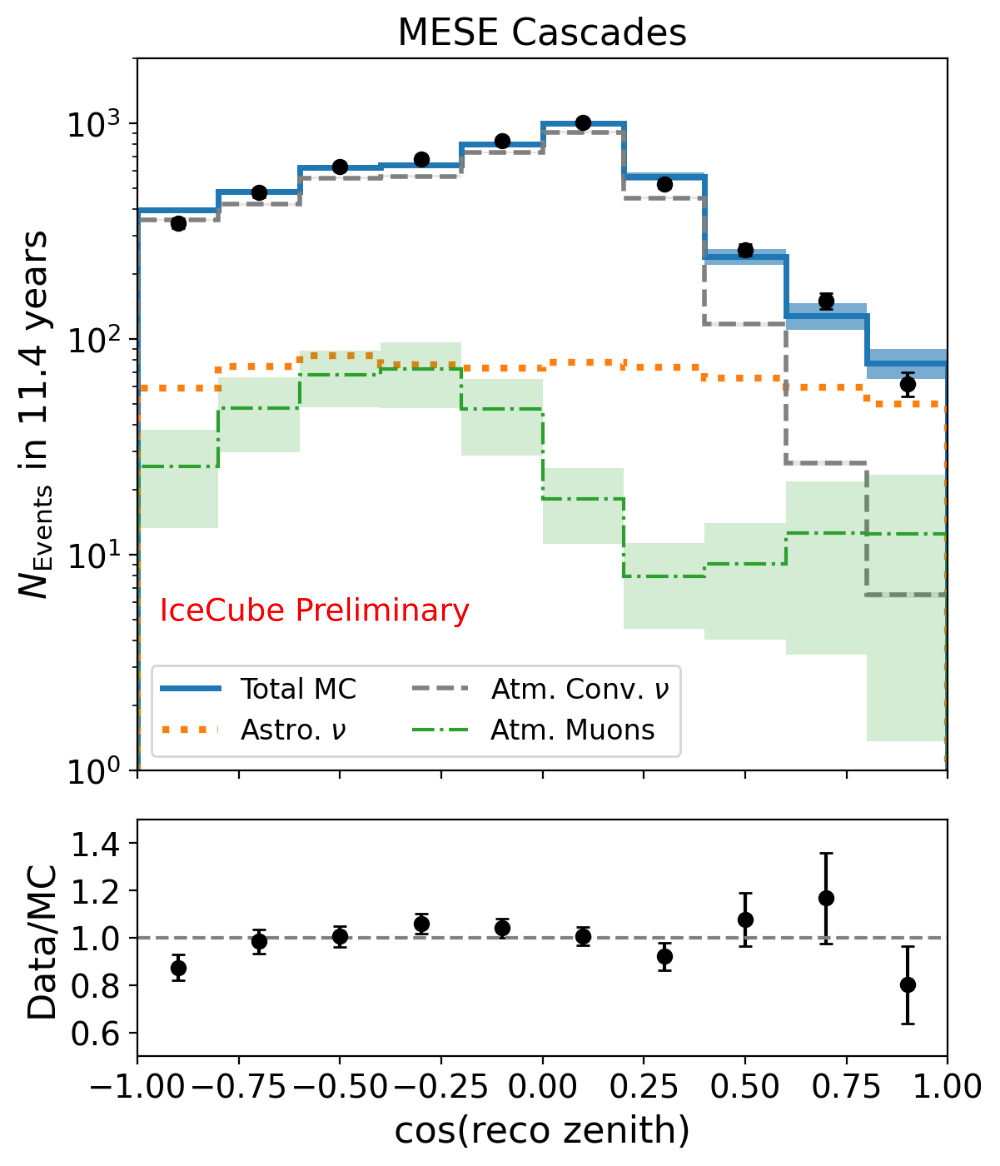}
\includegraphics[width=0.33\linewidth]{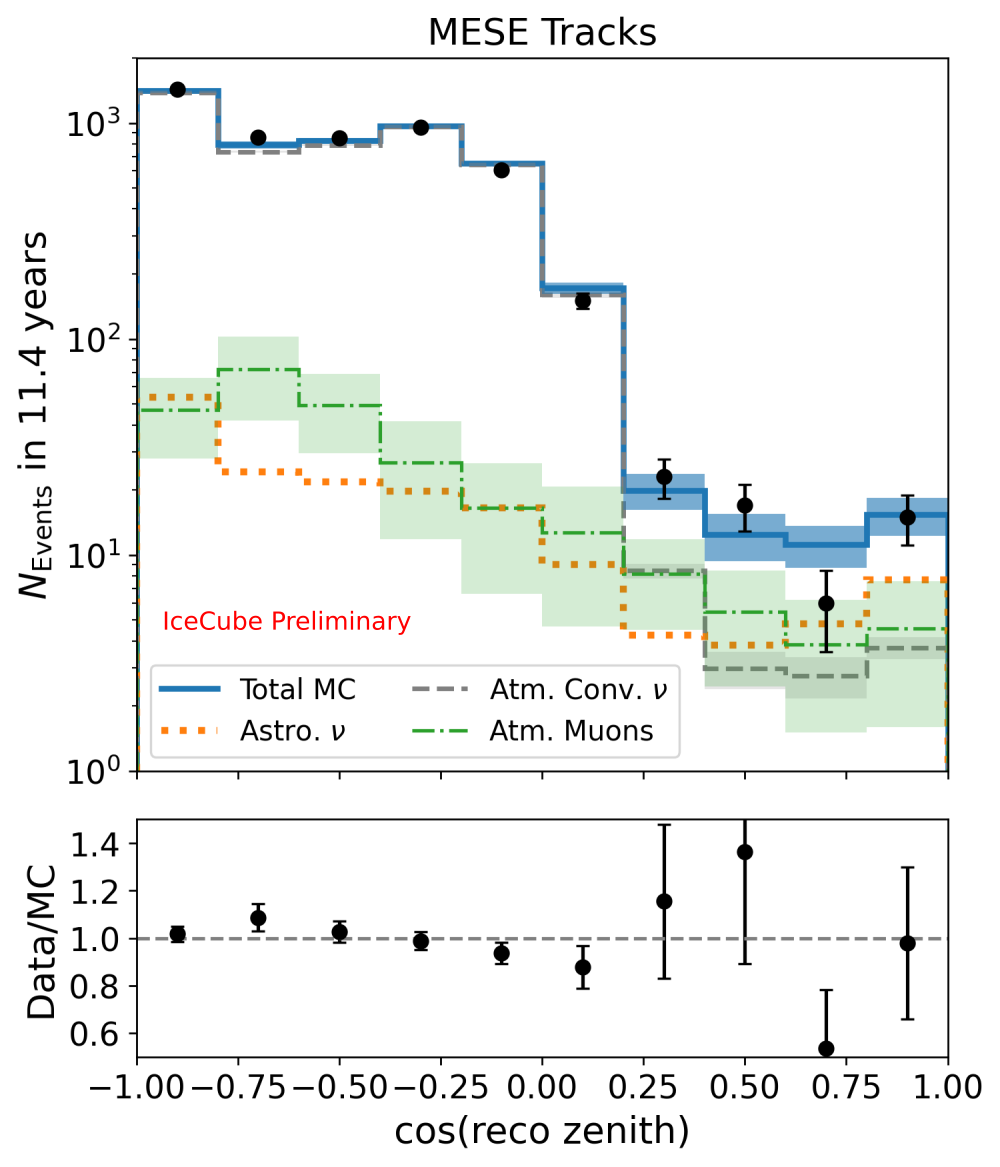}
\includegraphics[width=0.33\linewidth]{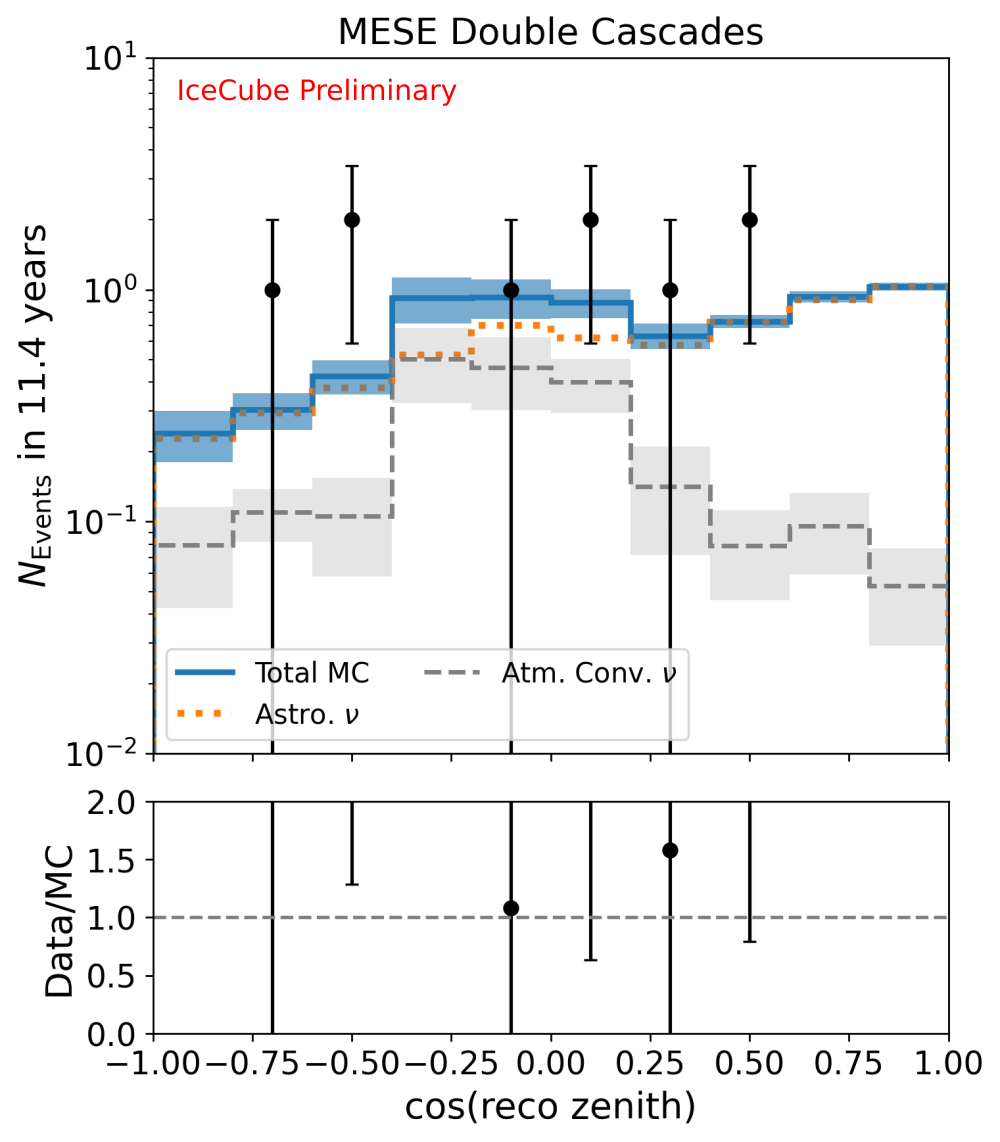}
\begin{center}
\includegraphics[width=0.33\linewidth]{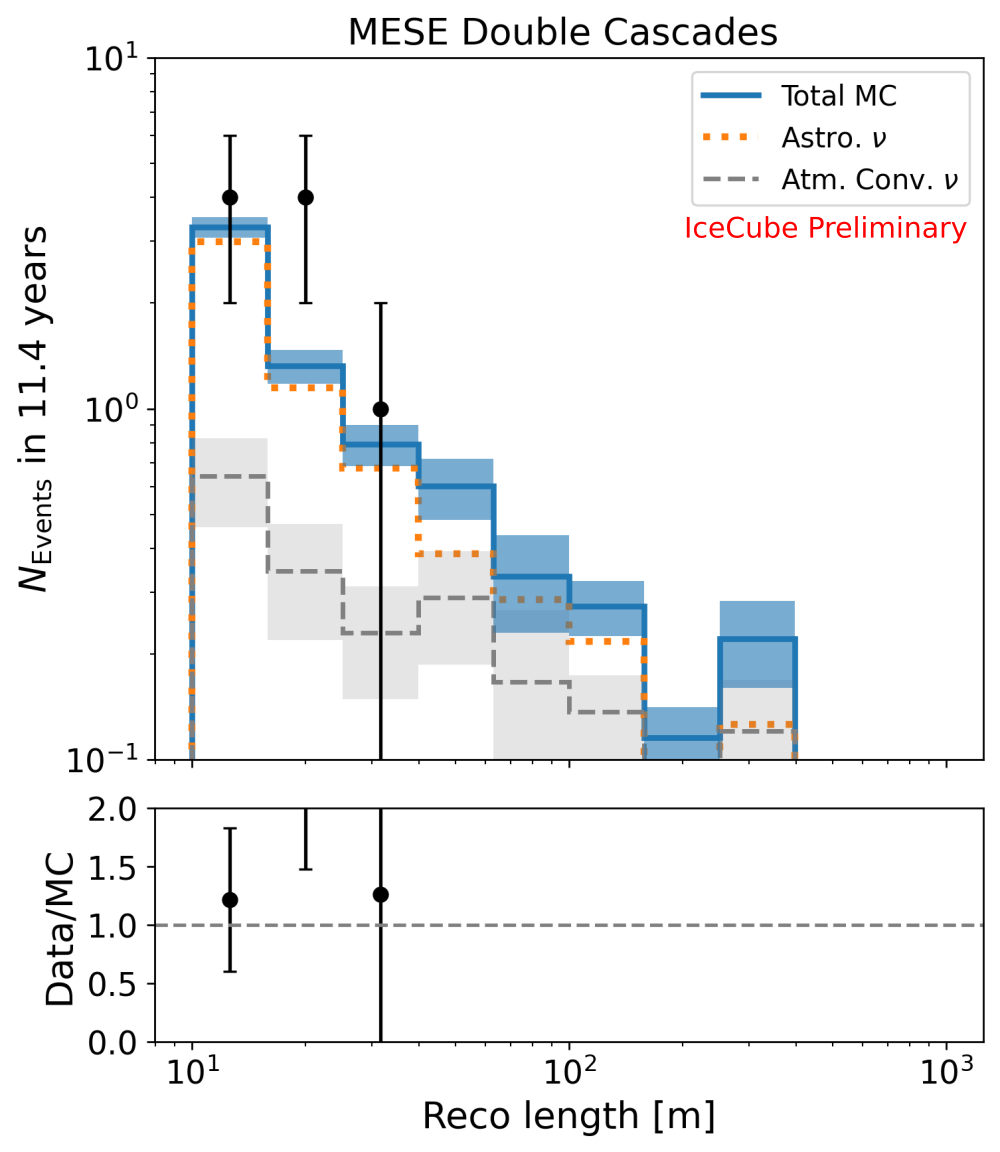}
\end{center}
\caption{The 1D projections of the reconstructed energy and reconstructed cosine zenith distributions of cascade, track, and double cascade classified events in MESE. The reconstructed length distribution for double cascade classified events is also shown. Data is shown in black and total best-fit MC is shown in blue. The individual components of the total MC are also shown in the figure. The prompt atmospheric flux fits to zero and is therefore not shown in the figure. No atmospheric muons pass the double-cascade classification.}\label{fig1}
\end{figure}

The energy distributions of cascades, tracks, and double cascades are shown in Figure~\ref{fig1}. Observed data is compared to the total MC, under our best-fit BPL spectrum with normalization $\phi\,=\,2.72^{+0.95}_{-0.92}\times\rm{10^{-18}/GeV/cm^2/s/sr}$, the lower energy spectral index $\gamma_1\,=\,1.76^{+0.36}_{-0.26}$, the higher energy index $\gamma_2\,=\,2.81^{+0.08}_{-0.12}$ and the break energy $\log_{10}(E_{\mathrm{break}}/\mathrm{GeV})\,=\,4.5^{+0.12}_{-0.09}$. The best fit astrophysical flavor ratio at Earth is obtained as $f_e:f_{\mu}:f_{\tau}\,=\,0.30 : 0.37 : 0.33$.
The figure shows that data and the best-fit MC are compatible with each other within 2\,$\sigma$. This is also the case for the other observables used in the fit: cosine of reconstructed zenith (for all three morphologies) and reconstructed length (for double cascades). 
Since only 9 double-cascade classified events exist in data, consistent with the expectation from MC, we have large statistical uncertanties in the double cascades histograms. 

\begin{figure}[h]
\centering
\includegraphics[width=0.6\linewidth]{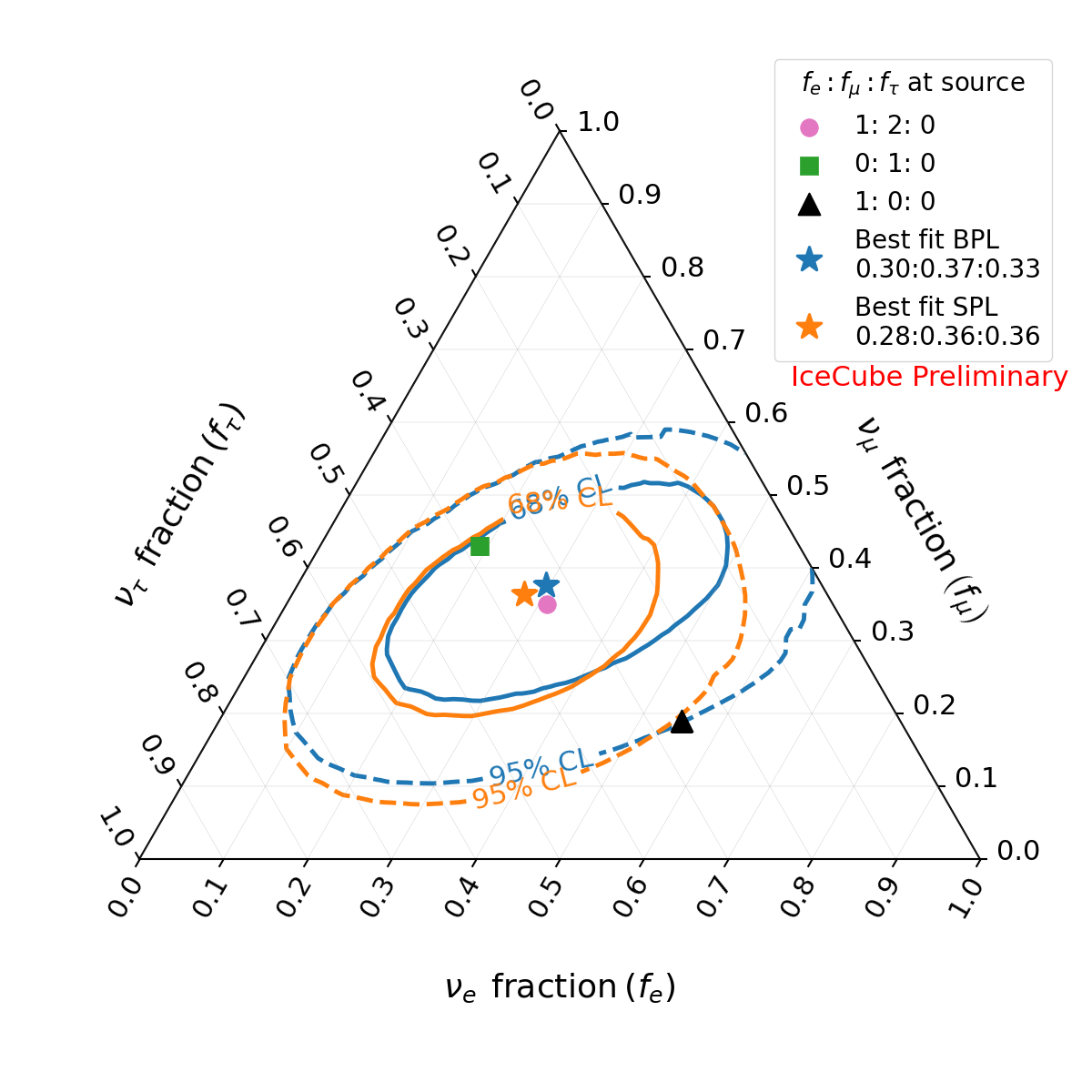}
\caption{Fits to the astrophysical flavor ratio with the MESE sample for the baseline assumption of a BPL flux (blue) and for the cross-check with an SPL fit (orange). 68\% CL and 95\% CL Wilks' contours are shown here along with the best fit. The best fit from both flux assumptions are consistent with each other. Assumptions of flavor ratio at source, after undergoing oscillations during their propagation towards Earth are also shown in the figure.}\label{fig2}
\end{figure}

Figure~\ref{fig2} shows the 68\% and 95\% Wilks' contours obtained from the fit on a ternary diagram showing the fraction of each neutrino flavor on Earth. Also shown in the figure are the expectations for several flavor assumptions at source, after averaged oscillations during their journey from the source to the Earth. These source assumptions are: pion decay resulting in a flavor ratio of 1:\,2:\,0 at source, muon damped pion decay with a ratio of 0:\,1:\,0, and neutron decay with a ratio of 1:\,0:\,0. The best fit is seen to be consistent with expectations from the standard theory of neutrino oscillations, as any deviation from this would result in the best fit lying in a region that is not along the line connecting the three standard source scenarios shown in the figure. 

The figure also shows the best fit and the Wilks' contours under the SPL assumption. While the best fits and the shape of the contours with the BPL and SPL fits remain comparable, it is seen that the 95\% CL contour closes under the SPL assumption and does not under the BPL assumption. 
The reason for this is the harder spectral index ($\gamma = 2.54^{+0.05}_{-0.04}$) fitted for the SPL model, resulting in a prediction of a larger number of high-energy neutrinos and, therefore, a larger fraction of events with longer tau decay-lengths, recognizable as double-cascades.
This results in the inability to close the 95\% contour for the BPL model. This also demonstrates the necessity to model the spectrum properly to obtain an accurate measurement of the flavor ratio.

\begin{figure}[h]
\centering
\includegraphics[width=0.6\linewidth]{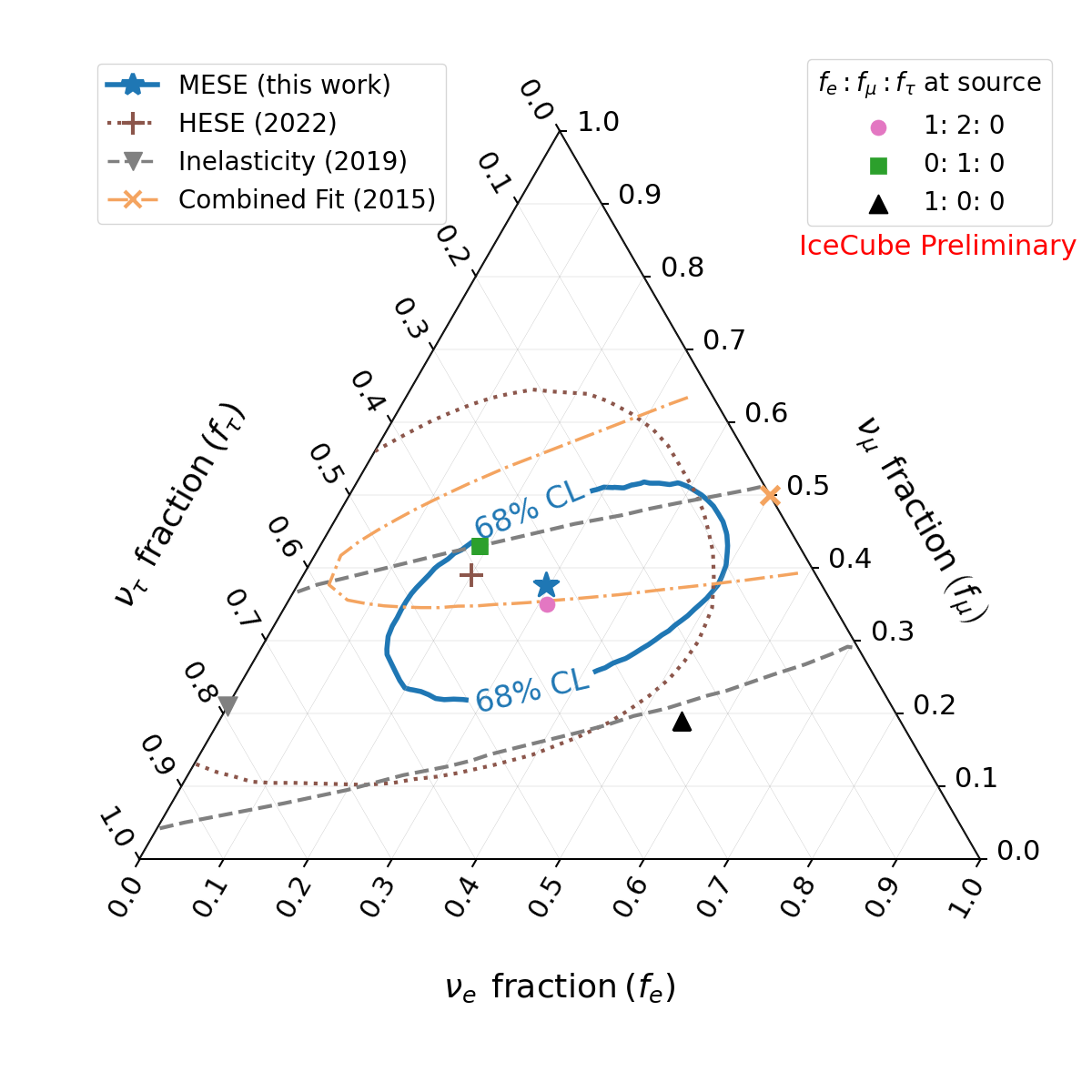}
\caption{A comparison of the flavor ratio measured with the MESE sample to previous measurements made with IceCube~\cite{icecube_collaboration_combined_2015}~\cite{IceCube:2018pgc}~\cite{IceCube:2020fpi} is shown. All curves show the 68\% CL Wilks' contours, along with the respective best fit. Previous measurements with IceCube assumed an SPL flux when measuring the flavor ratio, while this work assumes the flux to be a BPL, consistent with IceCube's latest spectral measurements~\cite{MESEdiffuse:2025icrc}. Including the double-cascades classification along with the cascades and tracks classification helps the contour close in the MESE analysis.}\label{fig3}
\end{figure}
Figure~\ref{fig3} shows the measurement of the flavor ratio in this analysis compared to previous measurements from IceCube. We can see that the 68\% CL contour closes for the first time with this analysis, marking a milestone in IceCube's measurements of the astrophysical flavor ratio. This can be attributed to the following:
\begin{itemize}
   \item Inclusion of tau-neutrino classification: This breaks the degeneracy between electron and tau neutrinos. For example, 'Combined Fit (2015)'~\cite{icecube_collaboration_combined_2015} and 'Inelasticity (2019)'~\cite{IceCube:2018pgc} in the figure did not include any classification of tau neutrino based morphologies. This makes it difficult to close the contours either along the $\nu_e$ or $\nu_{\tau}$ axis.
   \item Inclusion of TeV-scale events: This increases the total number of events, especially for the cascade and track morphologies when compared to using only the high energy events (above 60\,TeV) as done in 'HESE (2022)'~\cite{IceCube:2020fpi}. This in turn, gives strong constraints from higher statistics, especially along the electron and muon neutrino axes along with improved constraints on the nuisance parameters in the fit.
    \item An increased livetime: We use 11.4 years of data collected with IceCube in this analysis.
\end{itemize}
\section{Conclusion}\label{sec3}
\begin{justify}
The results presented here show that, for the first time, we are able to constrain the fraction of neutrinos of each flavor of neutrino to be $>0$ with more than 68\% CL. Based on the maximum-likelihood test, we reject zero electron neutrinos with 98.7\% CL and a zero fraction of tau neutrinos with 91.9\% CL. While the best fit flavor ratio of $f_e:f_{\mu}:f_{\tau}\,=\,0.30 : 0.37 : 0.33$ is closest to the standard pion decay scenario, the muon-damped source scenario is still within the 68\% CL contour. 
A neutrino flux dominated by neutron decay at source is rejected at 94.8\% CL with the Wilks' contour from the maximum-likelihood fit. 
Although a previous analysis from IceCube ('Combined Fit (2015)'~\cite{icecube_collaboration_combined_2015}) rejected the neutron-decay scenario at a higher confidence level, the previous analysis used no information about tau neutrinos. This analysis uses an updated treatment of systematics and better modeling of the ice, which affects light propagation and therefore the description of signal collection by the DOMs. 
The best fit of the flavor ratio lies on the line connecting the three standard source scenarios, which is the only region allowed by the standard theory of neutrino oscillations~\cite{Song:2020nfh}. Therefore, our results are consistent with this theory. Future 
studies including other methods of classifying tau-neutrino events, for instance a neural-network based classification~\cite{IceCube:2024nhk} or identification of tracks induced by tau neutrinos~\citep{IceCube:2018pgc,dissertationRClark} can be expected to provide a larger sample of tau neutrino classified events to be included in the flavor measurement. 
\end{justify}
\bibliographystyle{ICRC}
\bibliography{references}

%

\clearpage

\section*{Full Author List: IceCube Collaboration}

\scriptsize
\noindent
R. Abbasi$^{16}$,
M. Ackermann$^{63}$,
J. Adams$^{17}$,
S. K. Agarwalla$^{39,\: {\rm a}}$,
J. A. Aguilar$^{10}$,
M. Ahlers$^{21}$,
J.M. Alameddine$^{22}$,
S. Ali$^{35}$,
N. M. Amin$^{43}$,
K. Andeen$^{41}$
% ...
\\
\\
% ...
$^{10}$ Universit{\'e} Libre de Bruxelles, Science Faculty CP230, B-1050 Brussels, Belgium \\
% ...
$^{16}$ Department of Physics, Loyola University Chicago, Chicago, IL 60660, USA \\
$^{17}$ Dept. of Physics and Astronomy, University of Canterbury, Private Bag 4800, Christchurch, New Zealand \\
% ...
$^{21}$ Niels Bohr Institute, University of Copenhagen, DK-2100 Copenhagen, Denmark \\
$^{22}$ Dept. of Physics, TU Dortmund University, D-44221 Dortmund, Germany \\
% ...
$^{35}$ Dept. of Physics and Astronomy, University of Kansas, Lawrence, KS 66045, USA \\
% ...
$^{39}$ Dept. of Physics and Wisconsin IceCube Particle Astrophysics Center, University of Wisconsin{\textemdash}Madison, Madison, WI 53706, USA \\
% ...
$^{41}$ Department of Physics, Marquette University, Milwaukee, WI 53201, USA \\
% ...
$^{43}$ Bartol Research Institute and Dept. of Physics and Astronomy, University of Delaware, Newark, DE 19716, USA \\
% ...
$^{63}$ Deutsches Elektronen-Synchrotron DESY, Platanenallee 6, D-15738 Zeuthen, Germany \\
$^{\rm a}$ also at Institute of Physics, Sachivalaya Marg, Sainik School Post, Bhubaneswar 751005, India \\
% ...

\subsection*{Acknowledgments}

\noindent
The authors gratefully acknowledge the support from the following agencies and institutions:
USA {\textendash} U.S. National Science Foundation-Office of Polar Programs,
U.S. National Science Foundation-Physics Division,
U.S. National Science Foundation-EPSCoR,
U.S. National Science Foundation-Office of Advanced Cyberinfrastructure,
Wisconsin Alumni Research Foundation,
Center for High Throughput Computing (CHTC) at the University of Wisconsin{\textendash}Madison,
Open Science Grid (OSG),
Partnership to Advance Throughput Computing (PATh),
Advanced Cyberinfrastructure Coordination Ecosystem: Services {\&} Support (ACCESS),
Frontera and Ranch computing project at the Texas Advanced Computing Center,
U.S. Department of Energy-National Energy Research Scientific Computing Center,
Particle astrophysics research computing center at the University of Maryland,
Institute for Cyber-Enabled Research at Michigan State University,
Astroparticle physics computational facility at Marquette University,
NVIDIA Corporation,
and Google Cloud Platform;
Belgium {\textendash} Funds for Scientific Research (FRS-FNRS and FWO),
FWO Odysseus and Big Science programmes,
and Belgian Federal Science Policy Office (Belspo);
Germany {\textendash} Bundesministerium f{\"u}r Forschung, Technologie und Raumfahrt (BMFTR),
Deutsche Forschungsgemeinschaft (DFG),
Helmholtz Alliance for Astroparticle Physics (HAP),
Initiative and Networking Fund of the Helmholtz Association,
Deutsches Elektronen Synchrotron (DESY),
and High Performance Computing cluster of the RWTH Aachen;
Sweden {\textendash} Swedish Research Council,
Swedish Polar Research Secretariat,
Swedish National Infrastructure for Computing (SNIC),
and Knut and Alice Wallenberg Foundation;
European Union {\textendash} EGI Advanced Computing for research;
Australia {\textendash} Australian Research Council;
Canada {\textendash} Natural Sciences and Engineering Research Council of Canada,
Calcul Qu{\'e}bec, Compute Ontario, Canada Foundation for Innovation, WestGrid, and Digital Research Alliance of Canada;
Denmark {\textendash} Villum Fonden, Carlsberg Foundation, and European Commission;
New Zealand {\textendash} Marsden Fund;
Japan {\textendash} Japan Society for Promotion of Science (JSPS)
and Institute for Global Prominent Research (IGPR) of Chiba University;
Korea {\textendash} National Research Foundation of Korea (NRF);
Switzerland {\textendash} Swiss National Science Foundation (SNSF).

\end{document}